\documentclass{article}

\usepackage{graphicx}
\usepackage{amsmath}
\usepackage{amssymb}
\usepackage{natbib}
\usepackage[utf8]{inputenc}
\usepackage{tikz}

\newcommand{\Hessian}{\mathrm{He}}
\newcommand{\Ito}{It{\^o}}

\newcommand{\RR}{{\mathbf{R}}}
\newcommand{\qed}{\rule{2mm}{2mm}}
\newcommand{\transpose}{^\top}
\newcommand{\delvis}[2]{\frac{\partial #1}{\partial #2}}

\newcounter{envcounter} 
\newcommand{\myenv}[1]{\par\refstepcounter{envcounter}{\bf #1
    \theenvcounter:} }
\newenvironment{bemark}{\myenv{Remark}}{\mbox{ }\hfill$\qed$\par}
\newenvironment{lemma}{\myenv{Lemma}}{\mbox{ }\hfill$\qed$\par}

\begin{document}

\title{Inference in stochastic differential equations using the Laplace approximation: Demonstration and examples}
\author{Uffe Høgsbro Thygesen$^1$, Kasper Kristensen$^1$}
\date{$^1$ DTU Compute, \texttt{uhth@dtu.dk}\\
  $^2$ DTU Aqua, \texttt{kaskr@dtu.dk}\\
  Technical University of Denmark\\
  DK-2800 Kongens Lyngby, Denmark\\
  \today}
\maketitle

\abstract{Stochastic differential equations are a natural framework for dynamic systems and time series in ecology, because they allow for non-linear first-principle knowledge and uncertainty in the dynamics, and can be combined with measurement errors. However, estimation methods are often technically and computationally challenging. Here, we demonstrate that the Laplace approximation is useful for estimating states and parameters in these models, when done correctly. We give special attention to non-linear dynamics, state-dependent noise intensities, and non-Gaussian measurement errors. Our technique adds states between times of observations, approximates transition densities using discretization methods - in the simplest case, the Euler-Maruyama method - and eliminates unobserved states using the Laplace approximation. We demonstrate that consistency requires a particular form of the approximation, and provide different approaches to implementation. Using simulated case studies, we demonstrate that transition probabilities are well approximated, that inference is computationally feasible, and that the framework leads to simple and flexible implementations. }

\maketitle


\section{Introduction}
\label{sec:intro}

A common problem in statistical analysis of ecological time series is to combine mechanistic understanding of the system at hand with the ubiquitous presence of uncertainty, which together means that systems are only partially predictable. In all but the simplest cases, mechanistic ecological models contain non-linearities and lead to non-standard statistical distributions, so that the modeler must trade off fidelity to first principles with availability of statistical methods for analysis.

A conceptually appealing framework for ecological time series is state-space models \citep{Auger-Methe2021,Nabeel2025} based on stochastic differential equations \citep{Oeksendal2010,Thygesen2023sde}; see section \ref{sec:problem} for a detailed model specification. These models can include ecological mechanisms, external inputs, and unknown parameters, while the noise leads to uncertainty in the evolution of the system. Observation equations complement the dynamics by describing how measurements relate to states as well as the statistical properties of measurements. The continuous-time setting allows including mechanistic theory formulated as differential equations, and is convenient when the observations are irregular in time.

Although stochastic differential equations may appeal to a theoretical modeler, the technicalities of the analysis can be substantial, and several approaches co-exist \citep{Fuchs2013,Nabeel2025}.  An important special case is that of linear systems driven by Gaussian noise with constant intensity, and where linear combinations of the state variables are measured subject to Gaussian errors. Then the entire model is Gaussian, and the celebrated Kalman-Bucy filter \citep{Kalman1961,Simon2006} provides state estimates for given parameters. The predictive Kalman filter gives residuals which can be used for estimation of system parameters; maximizing the likelihood over parameters corresponds to tuning the one-step predictive filter \citep{Madsen2007}. The prediction residuals are also central to model validation \citep{Holst2003,Thygesen2017}. The Kalman filter can be applied to very high-dimensional systems; even more so with the extension of the Ensemble Kalman Filter \citep{Evensen2003}, which allows for spatio-temporal models common in the geosciences. Examples of ecological  applications of Kalman filters (in continuous or discrete time) include analysis of animal motion and population dynamics \citep{Lavender2025,Auger-Methe2021}.

Beyond the linear-Gaussian setting of Kalman filters, the problem of estimation in stochastic differential equations has a long history  within mathematical statistics \citep{PrakasaRao1979,Kessler1997,Nielsen2000} and remains an active research area \citep{Pilipovic2024,Jamba2024}. Weak nonlinearities can be addressed using the Extended Kalman filter and its many variants \citep{Simon2006,Kristensen2004}. For stronger nonlinearities, where the posterior distributions are not well approximated with Gaussians, sequential Monte Carlo techniques such as the particle filter \citep{Gordon1993} apply. For low-dimensional state spaces, the partial differential equation that governs the posterior distribution can be solved numerically as described, for example, by \cite{Pedersen2008} and \cite{Thygesen2023sde}. Simpler special cases arise if all states are measured perfectly at discrete deterministic points in time -  see, e.g., \citep{Pilipovic2024} and the references therein - or if the noise intensity does not depend on the state. 

From the point of view of practical modeling, the many methods for inference in stochastic differential equations represent a challenge, because no single method is universally applicable, and considerable technical expertise is needed to select and apply the most appropriate method in a given situation. Moreover, changes to a model which are minor seen from an ecological point of view, can mean that the chosen estimation method becomes infeasible. 

The difficulty with these estimation problems, which explains the many specialized methods, is the many unobserved random variables, for which the joint probability distribution is not available in  closed form, but which must nevertheless be integrated out to yield posterior distributions and likelihoods. The different methods represent different ways to to approximate these integrands and integrals. This paper focuses on the Laplace approximation (see, e.g., \cite{Kristensen2016} and the references therein), where the integrand is approximated with a Gaussian bell, for which the integral is known analytically. Thus, it replaces integration over a high-dimensional Euclidean space $\RR ^N$  with maximization of the integrand and computing the determinant of a Hessian. These operations are feasible even in very high dimensional spaces when the Hessian is sparse, as in time series analysis. Laplace's method has become increasingly popular in ecological statistics over the past decades thanks to availability of strong software such as ADMB \citep{Fournier2012}, INLA \citep{Rue2009}, and (R)TMB \citep{Kristensen2016,RTMB}. The Laplace approximation has been at the core of models for geolocation of tagged fish \citep{Albertsen2015}, stock assessment in fisheries \citep{Nielsen2016SAM},  spatiotemporal models \citep{Krainski2019}, and in general non-linear mixed-effects modelling  \citep{Brooks2017}.

The objective of this paper is to apply the Laplace approximation to estimation problems in stochastic differential equations, as they appear in statistical analysis of ecological time series. What characterize these problems are multiple states, non-linear dynamics, and incomplete measurements in discrete time with statistics that are not necessarily Gaussian. Importantly, the driving noise in the dynamics may be non-additive, i.e., vary in intensity with the state. For example, the noise in population dynamics scale with the population size and vanishes when the population is extinct. The Laplace approximation has already been applied to estimation of states and parameters in stochastic differential equations with additive noise \citep{Karimi2014,Karimi2016}; see also \citep[section 10.8]{Thygesen2023sde} as well as the linear example included in the RTMB package \citep{RTMB}. One contribution of the present work is to allow state-dependent noise intensities.

In the context of time series analysis, the first step in the Laplace approximation is to identify the maximum point of the integrand, i.e., the most probable state trajectory given data. Such dynamic optimization problems appear elsewhere in time series analysis; for example, in the Viterbi algorithm \citep{Zucchini2009} for Hidden Markov Models, and even in smoothing splines. A potential pitfall, which appears to have not been addressed before in this context, is that this variational problem should be stated more precisely as finding the most probable realization of the driving noise (section \ref{sec:trans-dense-diff-bridge}). 

Next, Laplace's approximation requires the transition probabilities in the underlying Markov process, which rarely are available in closed form, at least when models are multivariate and based on mechanistic reasoning. There exists a large array of approximate numerical methods \citep{Kloeden1999,AitSahalia2002,Pilipovic2024}; here, we add extra intermediate computational time points between observations, and integrate the corresponding unobserved states out using the Laplace approximation. In this way, Laplace's method is the unifying computational tool in our approach. 

The contribution of the present paper is to demonstrate that the resulting framework is relatively straightforward, has attractive computational performance, and provide reasonable approximations to transition probabilities and estimates of states and parameters. One advantage of the method is that it leads to simple and flexible implementations. A more careful analysis of the method, including in particular a certain continuous-time limit, will be explored elsewhere \citep{Thygesen2025sdeB}. As we will return to in the discussion, we believe the framework is applicable to a large class of problems, and can be refined and extended to cover even more situations. 

\section{Problem formulation}
\label{sec:problem}

We consider state-space models which can be written as a vector stochastic differential equation \citep{Oeksendal2010,Thygesen2023sde}
\begin{equation}
  \label{eq:sde}
  dX_t = f(X_t,t,\theta) ~dt + g(X_t,t,\theta)  ~dB_t .
\end{equation}
Here, $X_t\in \RR^n$ is the state of the system at time $t$, $\theta$ is a vector containing unknown parameters, and $\{B_t\}$ is $n$-dimensional standard Brownian motion; the equation is to be understood in the \Ito\ sense for now; we turn to the Stratonovich interpretation later. The term $f~dt$ describes the drift, i.e., the instantaneous change in the expectation of $X_t$, which can depend on observed driving inputs through the time dependence $t$. In turn, the term $g~dB_t$ represents uncertainty. We specify the initial condition $X_0$ through its probability density function $\pi(x)$. We assume that the usual conditions for existence and uniqueness of strong solutions apply; see, for example,  \citep{Oeksendal2010,Thygesen2023sde} for background material. We require that $g(x)$ is invertible for each $x\in \RR^n$, and write $g_k$ for the columns of $g$ ($k=1,\ldots,n$).

Next, the model includes an observation process $\{Y_i : i = 1,\ldots, N\}$ taken at discrete deterministic time points $0 \leq t_1<t_2 < \cdots < t_N = T$, summarized in the likelihood functions
\[
  l(x ; y_i , i, \theta) : i=1,\ldots,N 
\]
which are the conditional probability densities of the measurements $Y_i$ at the observed values $y_i$ given states $X_{t_i}=x$ and parameters $\theta$. These measurements can be univariate or multivariate and follow continuous or discrete distributions, even if we refer to probability \emph{densities}; we only require that the likelihood functions $l$ are smooth functions of the state $x$. 

As is well known, the joint  probability density function of all random variables $X_0,X_{t_1},\ldots,X_T$, $Y_1,\ldots, Y_N$ is then
\begin{equation}
  \label{eq:joint-pdf}
  \psi(x_0,\ldots,x_N,y_1,\ldots,y_N,\theta) = \pi(x_0) \prod_{i=1}^N p(t_{i-1},x_{i-1},t_{i},x_{i},\theta) \cdot \prod_{i=1}^N  l(x_i ; y_i , i, \theta) 
\end{equation}
where $p(s,x,t,y,\theta)$ is the transition density of the stochastic differential equation~\eqref{eq:sde}, i.e., the probability density function of $X_t$ evaluated at $y$, conditional on $X_s=x$, for parameters $\theta$. The likelihood function of parameters $\theta$ is obtained by integrating out the unobserved states 
\[
  L(\theta) = \int_{\RR^{n(N+1)}} \psi(x_0,x_1,\ldots,x_N,y_1,\ldots,y_N,\theta) ~ dx_0 \cdots dx_N , 
\]
and the posterior distribution of these states is, for given parameters $\theta$
\begin{equation}
  \label{eq:pdf-X-given-Y}
  f_{X|Y}(x,y; \theta) =   \psi(x_0,x_1,\ldots,x_N,y_1,\ldots,y_N,\theta) / L(\theta)
\end{equation}
The problem we address in this paper is how to evaluate this likelihood $L(\theta)$ numerically, so that numerical optimization is feasible, and so that the normalization in \eqref{eq:pdf-X-given-Y} can be performed. As described in the introduction, this problem is in general non-trivial and has been the subject of considerable research for two reasons: First, the transition probabilities $p$ are in general not available in closed form. Second, the integral is high-dimensional. In this paper, we focus on the method of the Laplace approximation \citep{Kristensen2016} to address both issues. This approximation considers a general integral $I(\epsilon)$ over $\RR^d$
\begin{equation}
  \label{eq:I-def}
  I(\epsilon) = \int_{\RR^d} \kappa(x) ~\exp(-\epsilon^{-1} \gamma (x)) ~dx
\end{equation}
where $\gamma,\kappa: \RR^d \mapsto \RR$ are $C^2$ and $\epsilon>0$ is a scaling parameter, and relies on the asymptote
\begin{equation}
  \label{eq:Laplace}
  I(\epsilon) \approx \kappa(\hat x)  \exp(-\epsilon^{-1} \gamma(\hat x)) \cdot \left| H/2\pi \epsilon \right|^{-1/2}
\mbox{ as } \epsilon \rightarrow0. 
\end{equation}
Here, the symbol ``$\approx$'' indicates that the ratio between the two expressions converge to 1 as $\epsilon\rightarrow 0$. On the right side, $\gamma(\hat x) = \min_x \gamma(x)$, $H$ is the Hessian of $\gamma$ evaluated at $\hat x$, and $|\cdot|$ denotes determinant. The approximation requires that  $\hat x$ is a unique global minimum point of $\gamma$ and that $H$ is positive definite.  The Laplace approximation can be understood as ``volume equals base area times height'', where $|H/2\pi \epsilon|^{-1/2}$ serves as the base ``area'' (volume in $\RR^d$). The Laplace approximation is exact when $\gamma(x)$ is quadratic in $x$ and $\kappa(x)$ is constant, so that the integrand is a Gaussian bell.  Beyond this, the justification of the approximation is that the majority of the integral comes from a small neighborhood near $\hat x$ where the integrand resembles a Gaussian bell. 

\begin{bemark}
  \label{remark:ambiguity}
  Note that a given integrand $\kappa(x) \exp(-\gamma(x))$ can be factored into $\kappa(x)$ and $\exp(-\gamma(x))$ in different ways, and that these will lead to different Laplace approximations. When we refer to ``the'' Laplace approximation $\hat I$, it is to be understood that we take $\kappa\equiv 1$ and apply the approximation with $\epsilon=1$, i.e.,
  \begin{equation}
    \hat I = \exp(-\gamma(\hat x)) \cdot \left| H/2\pi \right|^{-1/2} .
    \label{eq:hatI}
  \end{equation}
  This is also the approximation that the \texttt{R} packages \texttt{TMB} and \texttt{RTMB} compute, when given a user code that implements the function $\gamma$ \citep{Kristensen2016}.
\end{bemark}

\section{Evaluation of transition densities}
\label{sec:trans-dense-diff-bridge}

In this section we address the key sub-problem of numerical approximation of the transition probability density $p$ in the expression \eqref{eq:joint-pdf}. For a simpler notation, we ignore the parameters and focus on the time-invariant case
\begin{equation}
  \label{eq:ito-sde}
  dX_t = f(X_t) ~dt + g(X_t)  ~dB_t , \quad X_s = x. 
\end{equation}
Then, we aim to evaluate the transition probability density $p(s,x,t,y)$. Although the transition densities for stochastic differential equations are generally unknown, we can approximate them using methods from numerical analysis of stochastic differential equations when the time increment $t-s$ is small \citep{Kloeden1999}. We first use the simplest of these methods, viz., the Euler-Maruyama method
\begin{equation}
  \label{eq:E-M}
  X_{t + h } = X_{t} + f(X_{t}) ~ h  + g(X_t) ~( B_{t+h} - B_t), 
\end{equation}
so that, for suitably small time steps $h=t-s$, we approximate the transition densities with Gaussians
\begin{equation}
  \label{eq:EM-transprob}
  \hat p(s,x,t,y,h) = | \Sigma | ^{-1/2} \phi \left( \Sigma^{-1/2} (y-\mu) \right) . 
\end{equation}
Here, $\hat p(\cdot,h)$ indicates an approximation based on the time step $h$. We have used the shorthands $\mu = x + f(x) \cdot (t-s)$, $\Sigma = g(x) g\transpose(x) \cdot (t-s)$ for the conditional expectation and variance of $X_t$ given $X_s=x$, and throughout, we use $\phi(\cdot)$ for the p.d.f. of a multivariate standard Gaussian random variable, in any number of dimensions:
\[
  \phi(z) = | 2 \pi|^{-n/2} \exp( -|z|^2 / 2 ) \mbox{ for } z \in \RR^n .
\]
For our original problem of time series analysis, if there is ``fast sampling'' so that the time $t_{i} - t_{i-1}$ between observations is small compared to system dynamics, we can apply the Euler-Maruyama approximation directly. But when the sampling is not fast, we insert a number of intermediate computational time points $s=t_0 < t_1 < \cdots < t_N = t$. For given initial condition $X_s=x_0$, the Markov property of $\{X_t\}$ implies that the joint density of $X_{t_1},\ldots, X_{t_{N}}$ at $x_1,\ldots,x_N$ is
\begin{equation}
  \label{eq:bridge-FDD}
  \prod_{i=0}^{N-1} p(t_{i},x_{i},t_{i+1},x_{i+1}) . 
\end{equation}
The transition probability $p(s,x,t,y)$, i.e., the p.d.f. of $X_t = X_{t_N}$ at $x_N=y$, can be recovered from this expression by integrating out the intermediate variables $X_{t_1},\ldots,X_{t_{N-1}}$:
\begin{equation}
  \label{eq:bridge-integral}
  p(s,x,t,y) = \int_{\RR^{n(N-1)}}   \prod_{i=0}^{N-1} p(t_{i},x_{i},t_{i+1},x_{i+1})   \prod_{i=1}^{N-1} dx_{i} 
\end{equation}
where we take $x_0=x$ and $x_N=y$. The approach which is central to this paper is to approximate the short-term transition probabilities on the right hand side using discretization methods such as the Euler-Maruyama algorithm \eqref{eq:EM-transprob}, and employ the Laplace approximation to evaluate this high-dimensional integral. 

\subsection{The bias in the mode of the finite-dimensional distributions}
\label{sec:bias-mode-finite}

\begin{figure}
  \centering
  \includegraphics[width=0.5\textwidth]{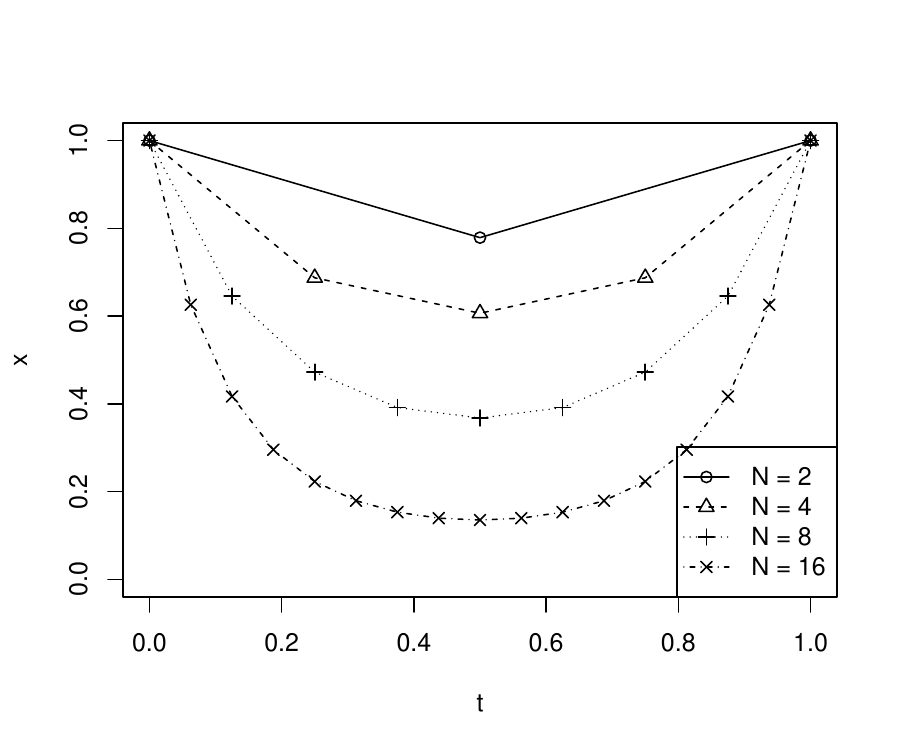}
  \caption{The Geometric Brownian Bridge: Modes of the finite-dimensional distributions for various numbers of interpolating points.}
  \label{fig:Geometric-Brownian-Bridge}
\end{figure}

In this section we demonstrate that a naive application of the Laplace approximation to the integral \eqref{eq:bridge-integral} leads to incorrect results, when the noise intensity $g(x)$ depends on the state $x$. 
For concreteness, we consider the scalar ($n=1$) example of geometric Brownian motion, which can for example be seen as a stochastic version of exponential growth from population dynamics \citep{Thygesen2023sde}. This process is given by the \Ito\ stochastic differential equation
\[
  dX_t = rX_t ~dt + \sigma X_t ~dB_t ,
\]
and has  log-normal transition densities, $X_t | X_s \sim LN( \log X_s + (r-\frac 12 \sigma^2) (t-s), \sigma^2 (t-s))$, when $X_s > 0$. We take parameters $r=1$, $\sigma=1$, and fix the initial and terminal condition, $s=0$, $X_s=x=1$, $t=1$, $X_t=y=1$; see figure \ref{fig:Geometric-Brownian-Bridge}. To reach a Laplace approximation of the transition probabilities based on \eqref{eq:bridge-integral}, we  insert a number $N-1$ of intermediate equally spaced time points, $t_i = i/N$, for $i=1,\ldots,N-1$. To eliminate errors arising from the Euler-Maruyama scheme, we use the true log-normal densities when evaluating the transition densities $p$ on the right hand side in \eqref{eq:bridge-integral} over the short time intervals. We then consider the diffusion bridge evaluated at those time points, and identify the mode of their joint density numerically. The result is seen in figure \ref{fig:Geometric-Brownian-Bridge}. Note that the mode in general lies below the line connecting the two end points, and does not show consistency as the number of intermediate time points are increased: The mode of $X_{u}$ converges to 0 as the number of intermediate points increases, for any $u\in (s,t)$. The implication is that the mode of the finite-dimensional distribution of the diffusion bridge should not be used as the basis of the Laplace approximation.  Indeed, a direct Laplace approximation of the integral (\ref{eq:bridge-integral}) leads to useless results, which are therefore not reported in detail.

The naive Laplace approximation of the integral \eqref{eq:bridge-integral} fails because the noise intensity $g(x)$ depends on the state variable $x$. Therefore, the state sequence can obtain high transition densities by moving to a region in state space where the noise intensity is low, i.e., near the origin. The shorter the time steps are, the more pronounced this bias will be, due to the scaling properties of diffusion, so that the mode of the finite dimensional distributions does not converge to a sensible bridge connecting the end points. Thus, an attempt to define a most probable realization of the diffusion bridge in this way will fail. Recalling that we, in general, need fine intermediate time steps for the Euler-Maruyama approximation \eqref{eq:EM-transprob} to be accurate, the Laplace approximation must shows consistency as the intermediate time steps vanishes. Therefore, we cannot base the Laplace approximation on the mode of the finite-dimensional distribution of the states. Although this example concerns computation of transition densities, the conclusion applies to the original estimation problem as well: The Laplace approximation should not be based on the sequence of states, which maximize the posterior probability density. 

\subsection{Laplace approximation in discretized Wiener space}
\label{sec:lapl-appr-wien}

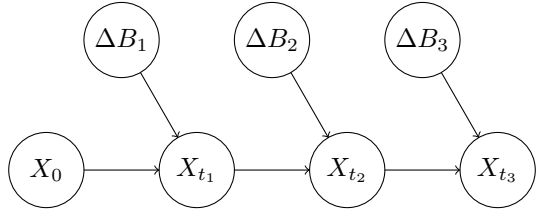
\begin{figure}
  \centering
  \begin{tikzpicture}
    \draw (0,0) circle(0.5) node {$X_0$};
    \draw[->] (0.5,0) -> (1.5,0);
    \draw (2,0) circle(0.5) node {$X_{t_1}$};
    \draw[->] (2.5,0) -> (3.5,0);
    \draw (4,0) circle(0.5) node {$X_{t_2}$};
    \draw[->] (4.5,0) -> (5.5,0);
    \draw (6,0) circle(0.5) node {$X_{t_3}$};
    
    \draw(1,1.73) circle(0.5) node {$\Delta B_1$};
    \draw(3,1.73) circle(0.5) node {$\Delta B_2$};
    \draw(5,1.73) circle(0.5) node {$\Delta B_3$};
    \draw[->] (1.25,1.3) -> (1.75,0.43);
    \draw[->] (3.25,1.3) -> (3.75,0.43);
    \draw[->] (5.25,1.3) -> (5.75,0.43);
  \end{tikzpicture}
  \caption{Probabilistic graphical network of random variables in a diffusion process $\{X_t\}$ driven by increments of the underlying Brownian motion $\{B_t\}$. }
  \label{fig:Markov-property}
\end{figure}

We now consider an alternative formulation, where we consider the Bayesian network in figure \ref{fig:Markov-property}. Here, the root random variables are the driving noise terms; specifically, the increments in the Brownian motion, $\Delta B_i = B_{t_i}- B_{t_{i-1}}$. From these, we can compute the intermediate states using the Euler-Maruyama scheme \eqref{eq:E-M}. Note that we can converting freely between states $X_{t_i}$ and increments $\Delta B_i$ in the Brownian motion, since the matrices $g(X_{t_i})$ remain invertible by assumption. This can be seen as a discretized version of the Laplace approximation in Wiener space as analyzed by \cite{Markussen2009}, so that a candidate continuous time limit exists - we will analyze this continuous-time limit elsewhere \citep{Thygesen2025sdeB}.

There are several ways to ensure that the path ends at $X_t=y$, which imposes a constraint on the increments in the Brownian motion. The perhaps simplest uses the following general and well known result:

\begin{lemma}
  \label{lemma:tiny}
  Let $Z\in \RR^m$ be a random variable with p.d.f. $f_Z(\cdot)$ and set $Y=g(Z)\in\RR^n$ where $g$ is continuous. Assume that the p.d.f. $f_Y(y)$ of $Y$ is well defined, then it equals
  \[
    f_Y(y) = \lim_{\epsilon\rightarrow0} \epsilon^{-n} \int_{\RR^m} f_Z(z) ~\phi(\epsilon^{-1}(y-g(z)))~dz . 
  \]
\end{lemma}

This lemma allows us to replace hard constraints between variables with soft constraints. 
To apply this to the situation of a constraint on the end point, let $\chi(b_1,\ldots,b_N)$ denote the end point $X_t$ of the trajectory as simulated with the Euler-Maruyama method \eqref{eq:E-M} using $\Delta B_i = b_i \in \RR^n$. Using lemma \ref{lemma:tiny}, we can find the p.d.f. of $X_t$ at $y$ as  
\begin{equation}
  \label{eq:B-integral}
\lim_{\epsilon\rightarrow 0} \int_{\RR^{Nn}}  \epsilon^{-n} \phi \left( (y- \chi(b_1,\ldots,b_N))/\epsilon \right) ~\prod_{i=1}^N \left( | \Delta t_i |^{-n/2} \phi\left(b_i/\sqrt{\Delta t_i} \right) ~db_i \right).
\end{equation}

We therefore pursue a Laplace approximation of this integral, for a fixed and finite but small value of $\epsilon$.

\subsection{Computational aspects}
\label{sec:comp}

The integral \eqref{eq:B-integral} may appear complicated, but it is straightforward to compute its Laplace approximation, for a given value of $\epsilon>0$. Here, we use the software \texttt{RTMB} (Template Model Builder for R), which is   available on CRAN; see also \citep{Kristensen2016}. This requires a function coded in \texttt{R} which evaluates the negative logarithm of the integrand for given values of $b_1,\ldots,b_N$, after which the software automatically maximizes the integrand, computes the Hessian at the maximum, and combines this to yield the Laplace approximation \eqref{eq:hatI}. For a scalar stochastic differential equation, the core user code consists of only a dozen lines, as is evident from the code that reproduces all results in this paper \citep{Thygesen_SDETMB}.

Although a direct implementation of the integral \eqref{eq:B-integral} is straightforward, it is inefficient from a computational point of view, so we do not detail the implementation (but see \cite{Thygesen_SDETMB}). The reason for this inefficienty is that the term $\chi(b_1,\ldots,b_n)$ causes the Hessian of the log-density to be non-sparse, which implies that evaluation of the determinant, needed for the Laplace approximation, is computationally expensive. The non-sparse component has low rank ($n$), so it is possible to address it analytically, but there are two more pragmatic approaches, as we will describe now.

\subsubsection{The "tiny" approach}
\label{sec:tiny-approach}

The first approach is to add a tiny random error to each time step in the Euler-Maruyama scheme, so that the Bayesian network in figure \ref{fig:Markov-property} is represented by
\begin{align*}
  \Delta B_i & \sim  N(0,I_n \cdot \Delta t_i), \\
  X_{t_i} | X_{t_{i-1}},\Delta B_i & \sim N( X_{t_i} + f(X_{t_{i-1}}) \Delta t_i + g(X_{t_{i-1}}) \Delta B_i,\epsilon^2 \Delta t_i I_n) . 
\end{align*}
This corresponds to adding an extra additive noise term, say, $\epsilon ~dW_t$ where $\{W_t\}$ is multivariate standard Brownian motion, to the original stochastic differential equation \eqref{eq:ito-sde}. Referring again to lemma \ref{lemma:tiny}, the original model is obtained in the limit $\epsilon\rightarrow0$, but we implement the algorithm with a small but finite $\epsilon$, say, $10^{-4}$, if other variables are on the order of unity. Now, latent random variables include both $\Delta B_i$ and $X_{t_i}$, and the Hessian of their joint density is sparse. This increases the dimension of the problem, but the computational cost of this is minor compared to the savings obtained with sparsity. The core code is roughly a dozen lines - see appendix \ref{sec:code} - while the entire code is available in \citep{Thygesen_SDETMB}.

\subsubsection{Formulation  in state space }
\label{sec:form-state-space}

Alternatively, we can eliminate the Brownian increments $\Delta B_i$ from the approximation \eqref{eq:B-integral} and perform the calculations in state space. With this, we obtain both the consistency from Wiener space, and the sparsity from state space. 

To see the effect of coordinate transformations on Laplace approximations, consider first the Laplace approximation \eqref{eq:hatI} of the integral of a general smooth function $\exp(-\gamma(x))$ as in \eqref{eq:I-def} with $\epsilon=1$ and  $\kappa(x) \equiv 1$. Now consider a smooth coordinate transformation $x=\eta(z)$ where $\eta$ is a diffeomorphism on $\RR^n$. From standard calculus, we have
\[
  \int_{\RR^n} e^{-\gamma(x)} ~dx = \int_{\RR^n} e^{- \gamma(\eta(z)) } ~ | \nabla \eta (z) | ~dz
\]
where $\nabla \eta$ is the Jacobian with elements $\partial x_i/\partial z_j$. Let $\hat I$ denote the Laplace approximation \eqref{eq:hatI} of the first integral, and define $\hat I_z$ as the Laplace approximation of the integral
\[
  \hat I_z \approx \int_{\RR^n} e^{- \zeta(z) }  ~dz ,
\]
where we have defined  $\zeta(z) = \gamma(\eta(z))$. This function $\zeta$ attains it global minimum at $\hat z=\eta^{-1}(\hat x)$, and the Hessian of $\zeta$ at this point is $H_z := \Hessian_z \zeta (\hat z) = J\transpose HJ$, where $J=\nabla \eta(\hat z)$ is the Jacobian of $\eta$ at $z$. It follows that
\[
  \hat I_z = \exp(- \zeta(\hat z)) | H_z / 2 \pi |^{-1/2} = \exp(- \gamma (\hat x)) | H / 2 \pi |^{-1/2} | J |^{-1} =  \hat I \cdot |J|^{-1}  . 
\]
Therefore, we can obtain the exact same Laplace approximation in either coordinate system, provided that we correct for the determinant $|J|$. This can also  be understood as different factorizations of the integrand in the two coordinate systems; compare remark \ref{remark:ambiguity}. 

To apply this general result for Laplace approximations in different coordinate systems to the problem of transition probabilities, we take  $\eta$ to be the coordinate transformation which maps the Brownian increments  $\{ \Delta B_i : i=1,\ldots,N \}$ to the states $\{ X_{t_i} : i=1,\ldots,N\}$ using the Euler-Maruyama scheme. Then,  the causality of system dynamics implies that the Jacobian $J = \nabla \eta $ is upper block triangular with block diagonal elements $g(X_{t_{i-1}})$, and thus the determinant is
\[
  | J | = \prod_{i=0}^{N-1} | g(X_{t_i}) | . 
\]
Note that $|J|\neq 0$, since we have assumed that $g(x)$ is (quadratic and) invertible for each $x$. An added benefit of this transformation is that the requirement that the path ends in $X_t=y$ is easier to express in state space than in the space of the Brownian increments, so we can resolve the limit $\epsilon \rightarrow 0$ in \eqref{eq:B-integral} analytically. The end result is the integral
\begin{equation}
  \label{eq:Laplace-state-space}
|J|^{-1} \int_{\RR^{(N-1) n }}  \prod_{i=1}^N \left[  |\Delta t_i|^{n/2} ~\phi \left( |\Delta t_i|^{-1/2} ~g^{-1}(x_{i-1}) \cdot ( x_i - x_{i-1} - f(x_{i-1}) \Delta t_i  \right)   \right]\prod_{i=1}^{N-1}  ~dx_i .
\end{equation}
Here, we still take $x_0=x$ and $x_N=y$. To compute the Laplace approximation of this integral, we define the vector of all intermediate points
$\bar x = (x_1,x_2,\ldots,x_{N-1})$, and 
\[
  \gamma(\bar x) = - \sum_{i=1}^N [ \log \phi( | \Delta t_i|^{-1/2}b_i )  - \frac n2 \log \Delta t_i ]
\]
where we have used the shorthand $b_i = g^{-1} (x_{i-1}) (x_i - x_{i-1} - f(x_{i-1}) \Delta t_i )$; this computes the Brownian increments that correspond to a given path in state space.   Then, the integral takes the form \eqref{eq:I-def} and we can use the the Laplace approximation \eqref{eq:hatI} to approximate the integral. To be explicit, let $\bar x^* = (x_1^*,x_2^*,\ldots,x_{N-1}^*)$ be the argument which minimizes $\gamma$; let $H$ be the Hessian of $\gamma$ at $\bar x^*$. Then the Laplace approximation is
\begin{equation}
  \label{eq:Laplace-state-space-final}
   | H / (2 \pi) |^{-1/2} \exp(-\gamma(\bar x^*)) \cdot \prod_{i=0}^{N-1} {|g(x^*_i)|^{-1}} . 
\end{equation}
Note that all terms in the log-integrand involve states at only two subsequent time points, so that the Hessian $H$ will be sparse; in the scalar case, tri-diagonal. To implement this expression   \eqref{eq:Laplace-state-space-final}
, we use the automated Laplace approximation in \texttt{RTMB} for the first two factors, and then correct the result for the last factor involving $|g(x^*_i)|$. The final code is roughly as complex or simple as the one discussed in section \ref{sec:tiny-approach} and presented in appendix \ref{sec:code}; see \cite{Thygesen_SDETMB}. 

\section{A Stratonovich formulation}
\label{sec:strat-form-heun}

\newcommand{\fS}{f_S}

While the previous has been for the \Ito\ formulation of stochastic differential equations, we now consider the Stratonovich stochastic differential equation 
\begin{equation}
  \label{eq:Strat-SDE}
  dX_t = \fS(X_t) ~dt + g(X_t) \circ dB_t
\end{equation}
where $\circ dB_t$ indicates the Stratonovich integral. Our motivation for including the Stratonovich interpretation in the analysis is not just completeness, but also that the continuous-time limit is more tractable in the Stratonovich case, as will be reported elsewhere \citep{Thygesen2025sdeB}. Recall \citep{Oeksendal2010,Thygesen2023sde} that if we choose
\[
  \fS(x) = f(x) - \frac 12 \sum_{k=1}^n \nabla g_k(x) g(x)
\]
then this Stratonovich equation \eqref{eq:Strat-SDE} governs (a version of) the same stochastic process $\{X_t:t\geq 0\}$ as the \Ito\ equation \eqref{eq:ito-sde}. The \Ito\ and Stratonovich equations can therefore describe the same phenomena, and is discussed by several authors \citep[for example]{Braumann2007}, the choice between the two formalisms is typically based on how the modeler perceives drift and noise, as well as on which formulation leads to the technically simplest analysis in the situation at hand. 

For the Stratonovich equation \eqref{eq:Strat-SDE}, the simplest time discretization is the implicit Euler-type, or trapozoidal, approximative relationship between states and Brownian increments:
\begin{equation}
  \label{eq:Strat-approximation}
  X_{t+h} - X_t = \frac 12 (\fS(X_{t}) + \fS(X_{t+h})) ~h  + \frac 12 (g(X_{t}) + g(X_{t+h})) ~(B_{t+h} - B_t) . 
\end{equation}
Note that the noise intensity $g(\cdot )$ is evaluated at both left and right end points of the time interval $[t,t+h]$, for consistency with the Stratonovich interpretation. To our knowledge, this equation is not being used to simulate sample paths of diffusion processes seen as initial value problems, because it is implicit in the new state $X_{t+h}$ and may lead to unbounded moments of $X_{t+h}$ conditional on $X_t$ \citep{Kloeden1999}.  However, it is convenient for the task of approximating transition densities between known points $X_t$ and $X_{t+h}$ when the time step $h$ is small. The details of this, and the Laplace approximation, are found in appendix \ref{app:Strat}. The user code consists of roughly a dozen lines for scalar equation \citep{Thygesen_SDETMB}.

\section{Transition densities in the Cox-Ingersoll-Ross process}
\label{sec:an-example:-cox}

To demonstrate the performance of our Laplace approximations, we consider the Cox-Ingersoll-Ross process, which is governed by the \Ito-sense stochastic differental equation
\[
  dX_t = \lambda(\xi - X_t) ~dt + \gamma \sqrt{X_t} ~dB_t ,
\]
or equivalently the Stratonovich equation
\[
  dX_t = \lambda (\xi - X_t) ~dt - \frac 14 \gamma^2 ~dt + \gamma \sqrt{X_t} \circ dB_t .
\]
All parameters $\lambda$, $\xi$ and $\gamma$ are positive, as is the initial condition $X_0=x$. This process can model population dynamics with outbursts; it was originally applied to interest rates \citep{Cox1985}. The model serves as a useful benchmark: The transition densities are well known \cite[exercise 9.8, for example]{Thygesen2023sde} and available in software \cite[for example]{Thygesen_SDEtools}. In stationarity, it is Gamma distributed with expectation $\xi$ and variance $\gamma^2 \xi / (2 \lambda)$; its autocorrelation decays exponetially with rate $\lambda$. The non-dimensional quantity $\lambda \xi/\gamma^2$ can be viewed as a signal-to-noise ratio. With weak noise, the process is similar to Ornstein-Uhlenbeck fluctuations around the expectation $\xi$, but for larger noise levels, the process becomes more skewed and intermittent. 

\begin{figure}
  \centering
  \includegraphics[width=\textwidth]{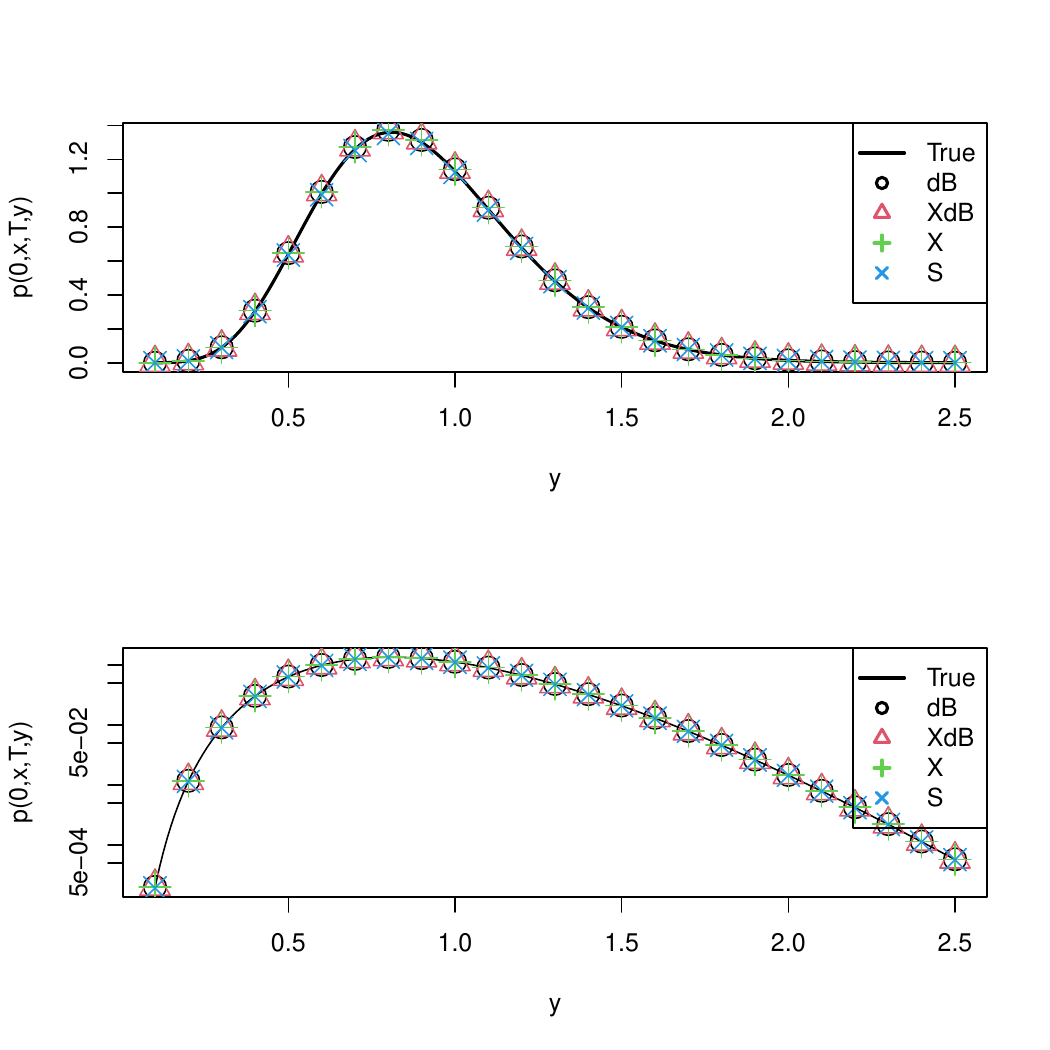}
  \caption{Transition probabilities for the CIR process in natural scale (top panel) and log scale (bottom panel): Comparison between the analytical densities (circles), and the densities computed using the Laplace approximation using the four different methods, which give practically indistinguishable results. See the text for parameters and full description. }
  \label{fig:CIR-transprob}
\end{figure}

We aim to approximate the transition densities $p(0,x,t,y)$ for various values of $y$, for $t=1$, $x=0.5$, and parameters $\lambda=1$, $\xi=1$, $\gamma = 0.5$. Note that the time $t$, which in the context of time series is the time between observations, is equal to the decorrelation time of the process; thus, an approximation of $p$ which is based on small $t$ is not likely to be very accurate. In stead, we approximate $p$ using the Laplace approximation and the four algorithms described in the previous: using the increments $B_{t_{i+1}} - B_{t_i}$ as latent variables (``dB''), using both states $X_{t_i}$ and increments and the ``tiny'' approach (``XdB''), using only states (``X''), and finally using the Stratonovich interpretation with only states (``S''). The results are shown in figure \ref{fig:CIR-transprob} and compared with the analytical result.

\begin{figure}
  \centering
  \includegraphics[width=\textwidth]{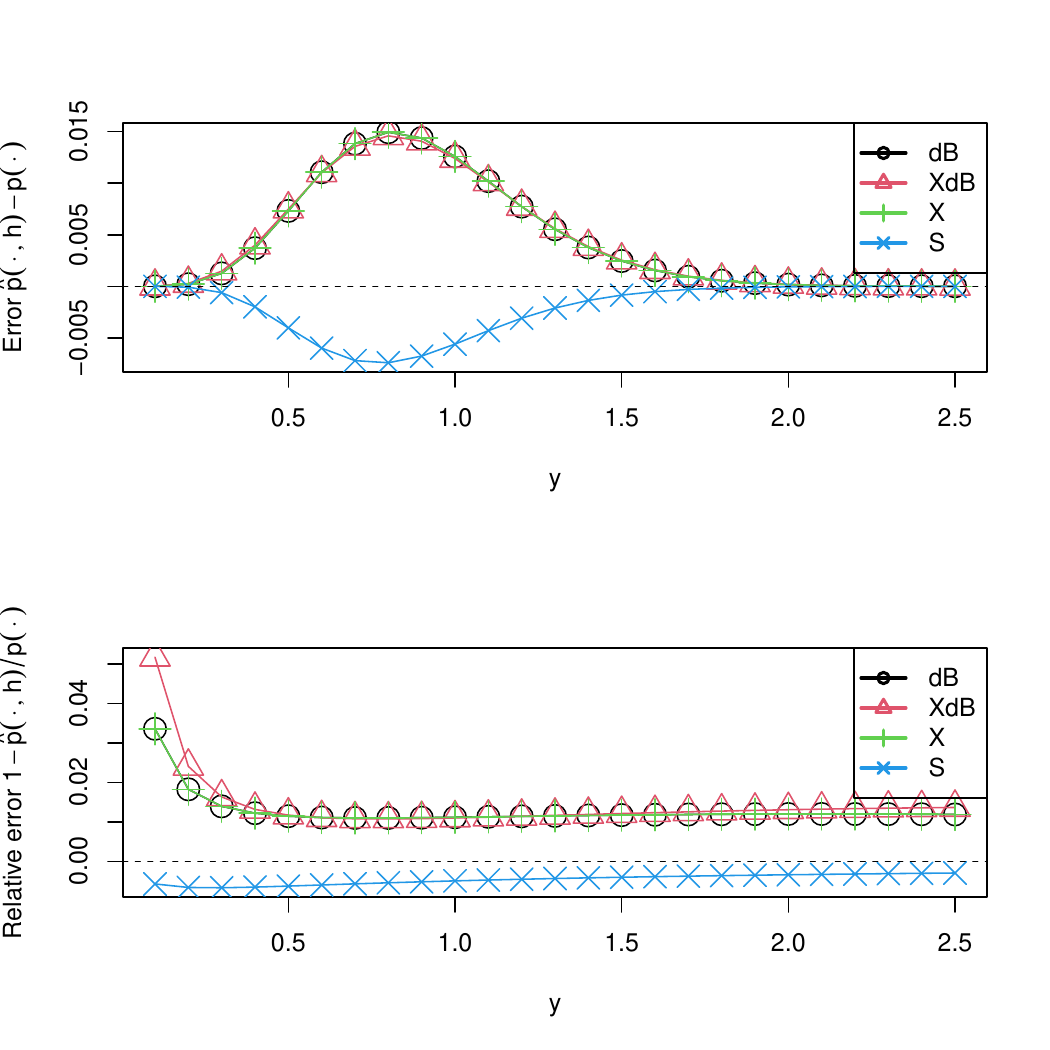}
  \caption{Errors in the transition probabilities for the CIR process: Absolute errors $\hat p-p$ (top panel) and relative errors $\hat p/p - 1$ (bottom panel), for the four different methods. The curves for the \Ito-based methods ``dB'' and``X'' are indistinguishable. See the text for parameters and full description. }
  \label{fig:CIR-transprob-err}
\end{figure}

We see that all methods lead to small approximation errors, both absolutely and relatively, and both near the mode and in the tails. It is not possible to discern the methods from each other in the graph. For that reason, figure \ref{fig:CIR-transprob-err} shows the error, both absolute and relative, between the approximations $\hat p(\cdot)$ and the analytical result $p(\cdot)$. For the absolute error, all methods lead to errors with somewhat similar magnitude, which are largest around the peak of the distribution. The methods ``dB'' and ``X'' lead to identical results up to machine precision, as predicted by the theory, while the method ``XdB'' leads to slightly different results, stemming from the parameter $\epsilon$ which broadens the distribution. For the relative error, the Stratonovich-based approximation leads to  lower errors in the tails of the distribution.

For these graphs, the computational time step was chosen excessively as $2^{-10}$, so that discretization does not affect the result. Computing times to compute the data in these graphs were 337 seconds, 0.4 seconds, 0.7 second and 1.1 seconds for the four methods, on a standard laptop. Note that the method ``dB'' based on the increments $\Delta B$ stands out as being significantly slower, due to the non-sparse Hessian; this method should not be used in practice, but serves only as a useful starting point for the development.

% Table of computing times in ../R/CIR-TP-times.txt

\begin{figure}
  \centering
  \includegraphics[width=\textwidth]{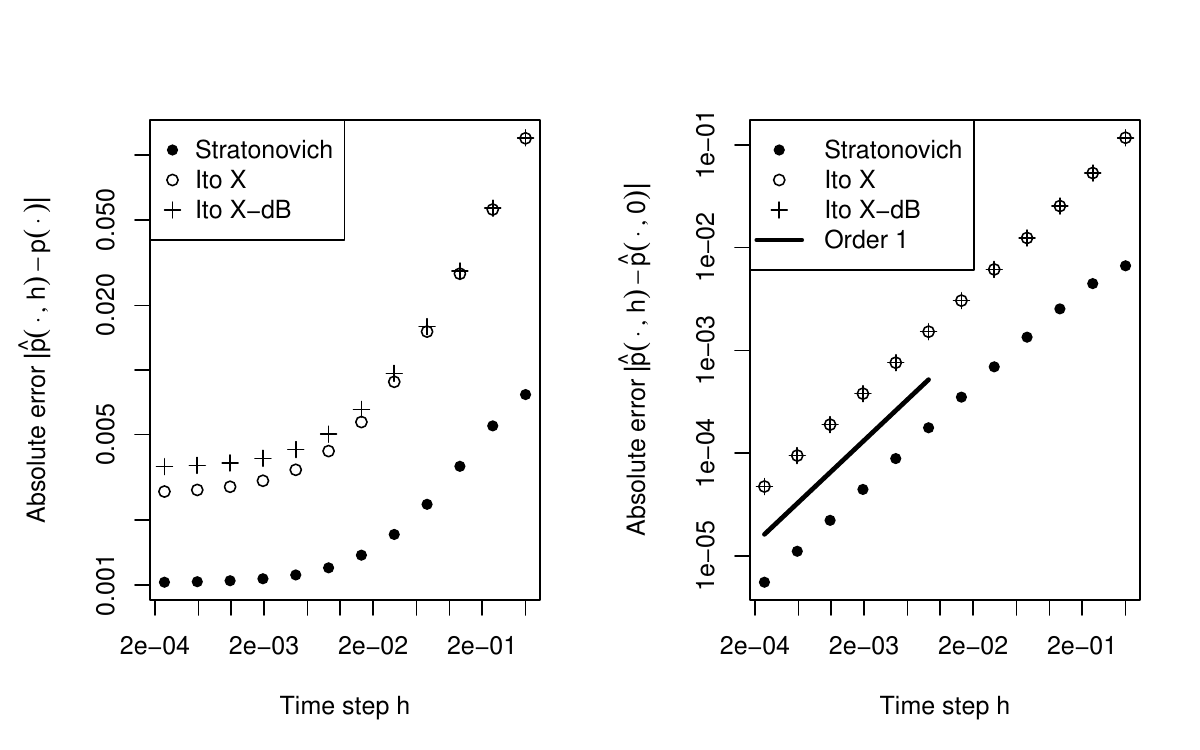}
  \caption{The effect of the computational time step on the error in the Laplace approximation of $p(0,3/4,1,3/2)$ for the CIR process.
    \emph{Left panel: } The total error against the true transition probability. 
    \emph{Right panel: } The error of time discretization, measured against the estimated continuous-time Laplace approximation. 
    Parameters as in figures   \protect\ref{fig:CIR-transprob}.}
  \label{fig:CIR-Timestep}
\end{figure}

The effect of the computational time step is seen in figure \ref{fig:CIR-Timestep}. As the time grids are refined, the errors do not vanish; an error due to the Laplace approximation itself remains, which depends on the method used (left panel). For coarser time grids, time discretization becomes more pronounced and the methods give more different results, with the Stratonovich-based method being most accurate. We can assess the discretization error by comparing with an  estimated limit of vanishing computational time steps obtained by Richardson extrapolation, specifically $\hat p(\cdot,0) = 2 \hat p(\cdot,h) - \hat p(\cdot,2h)$. Here, $h$ is the finest time step, and the expression assumes a linear relationship (order 1). The result is seen in the right panel. The plot confirms that the approximations are first order accurate. Note that the discretization error is identical for the two \Ito\ based methods, whereas the parameter $\epsilon$ introduces a fixed error, independent of the time step. We explore the continuous-time limit and the effects of time discretization in greater depth elsewhere \citep{Thygesen2025sdeB}.

\section{Inference in a predator-prey model}
\label{sec:R-MacA}

In this section we demonstrate by means of a simulation experiment that the approach applies not only to computation of transition densities, but also extends to a full time series analysis including estimation of both states and parameters. We consider a system of two  coupled stochastic differential equations, namely a stochastic predator-prey model of Rosenzweig-MacArthur type given by the two \Ito\ equations:

\begin{eqnarray}
  \label{eq:R-MacA-N}
  dN_t &=& rN_t(1-N_t/K) ~dt - C_t ~dt + \sigma_N N_t ~dB^{(1)}_t  - \sigma_C C_t ~dB^{(3)}_t, \\
  \label{eq:R-MacA-P}
  dP_t &=&  \epsilon C_t ~dt - \mu P_t ~dt + \sigma_P P_t ~dB^{(2)}_t + \sigma_C \epsilon C_t ~dB^{(3)}_t . 
\end{eqnarray}

Here, $N_t$ is the abundance of prey while $P_t$ is the abundance of the predators; $C_t = \beta N_t P_t / (1 + \beta N_t /\gamma ) $ is the consumption of prey by predators. Noise-free versions of these equations were examined by \cite{Rosenzweig1963}. Compared to the celebrated Lotka-Volterra system, it features logistic growth of prey in absence of predators, and a Holling type II functional response to represent saturation in the uptake per predator as the prey abundance increases \citep{Murray1989}. The stochastic version features environmental noise signals  $B^{(1)}_t,B^{(2)}_t$ which affect the two populations independently, with a noise intensity that scales with the abundances, and a third noise signal $B^{(3)}_t$ that represents multiplicative noise on the consumption process $C_t$. We note that several other stochastic versions can be envisioned, but our objective here is not to pursue this but simply demonstrate a time series analysis based on this model. Different aspects of a related stochastic model are discussed in some detail by \cite{Thygesen2023sde}. Parameters, their interpretation, and their values for the numerical experiment are given in table   \ref{tab:RmA-parameters}. With these parameters, the non-trivial equilibrium of the deterministic model is unstable and a stable limit cycle exists. 

\begin{table}
  \centering
  \begin{tabular}{cllrc}
    Symbol & Interpretation & True & Estimate &    std.dev
    \\
    \hline 
$r$       & Specific growth rate & 1   & 1.00    & 0.06 \\
$K$       & Carrying capacity & 1   & 1.06   & 0.10 \\
$\epsilon$ & Efficiency & 3   & -    & -  \\
$\beta$    & Clearance volume & 2   & 1.87    & 0.40  \\
$\gamma$    & Max. uptake per predator & 1   & -   & -  \\
$\mu$      & Pred. mortality & 1   & 0.96    & 0.12 \\
$\sigma_N$      & Prey noise  & 0.2 & 0.20   & 0.04\\
    $\sigma_P$      & Predator noise  & 0.1 & - & -   \\
    $\sigma_C$      & Consumption noise  & 0.1 & 0.0 & 0.08   \\
    $v$ & Sample volume & 100 & - & -  \\
    $h$ & Sample time & 1 & - & - 
  \end{tabular}
  \caption{Parameters in the stochastic Rosenzweig-MacArthur model.}
  \label{tab:RmA-parameters}
\end{table}

\begin{figure}
  \centering
  \includegraphics[width=\textwidth]{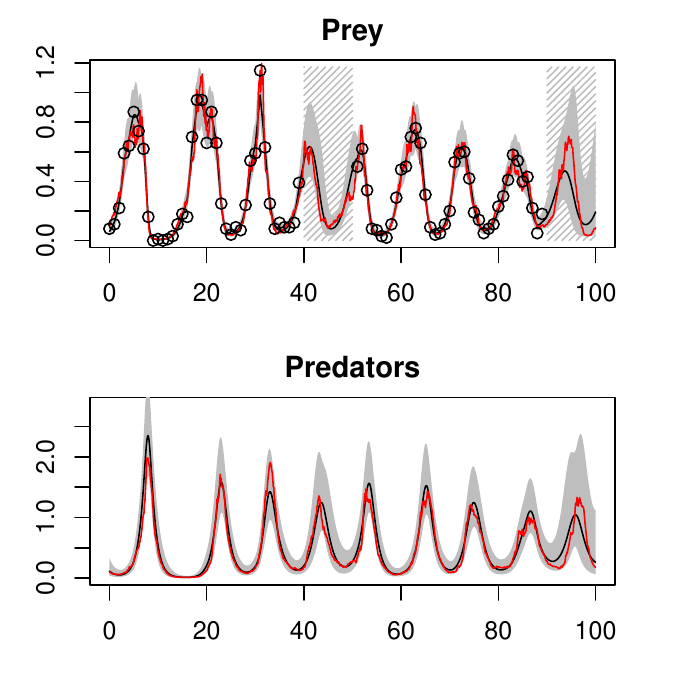}

  \caption{Simulated trajectory (solid red curve) and re-scaled measurements (circles) of the stochastic Rosenzweig-MacArthur model, along with re-estimated trajectory (solid black lines) and marginal 95 \% confidence intervals (grey region). }
  \label{fig:RmA-sim}
\end{figure}

A simulation of the system is seen in figure~\ref{fig:RmA-sim}. Here, we have included noisy measurements of the prey, generated as Poisson variables with expectation $v_N \cdot N_{t_{i}}$; the parameter $v_N=100$ can seen as a measure of sampling effort and determines the signal-to-noise level in the observations. Measurements are taken regularly every one time unit, while the computational time step is 0.1 time units. A period with missing data is included, and the simulation extends beyond the last observation to demonstrate forecasting abilities. 

Based on the simulated data set in the figure, the parameters in the model are re-estimated using Maximum Likelihood;  table \ref{tab:RmA-parameters} reports estimates as well as standard errors on the estimates based on the Hessian of the log-likelihood. Since the predators are not observed, the model is not completely parameter identifiable; specifically, the maximum consumption per predator $\gamma$ can be chosen freely to yield the same likelihood, provided that the abundance $P_t$, the clearance rate  $\beta$, and the conversion factor $\epsilon$  are scaled accordingly. Therefore, we fix the parameter $\gamma$. Even so, it is ambitious to estimate the details of the predator dynamics without observing predators, so we fix also the conversion efficiency $\epsilon$ and the noise level $\sigma_P$. The estimates of remaining parameters are computed using the ``XdB'' method, i.e., with the Brownian increments as root latent variables and the  ``tiny'' approach of section \ref{sec:tiny-approach}. Note that with this method, it is not problematic that the number of noise signals (3) exceeds the number of states (2). As latent variables, we choose the log-abundances $(\log N_t, \log P_t)$, since the abundances themselves are known to be positive. We compute the (Laplace approximation of) the posterior mean of the states, and marginal confidence regions, in log-domain and transform to original variables before plotting in figure \ref{fig:RmA-sim}. 

\begin{figure}
  \centering
  \includegraphics[width=\textwidth]{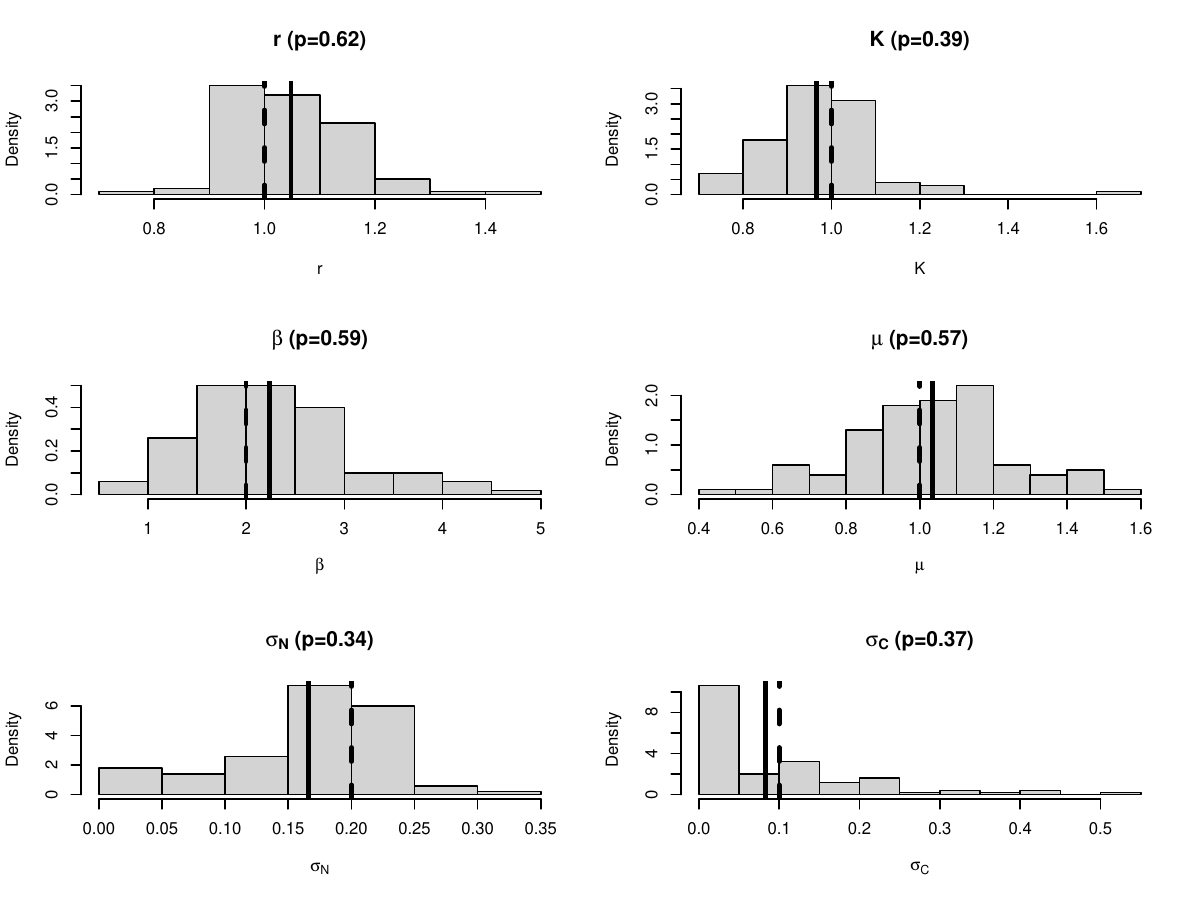}

  \caption{Histogram of parameter estimates, based on re-estimation from 100 simulated datasets. Included are the true values used for simulation (dashed) as well as the empirical mean of the estimates (solid). For each parameter, it is indicated what fraction of estimates are below the true value. }
  \label{fig:RmA-estimates}
\end{figure}

Next, we repeat the simulation-reestimation a total of 100 times. Figure \ref{fig:RmA-estimates} shows the distribution of the parameter estimates over the replicates. Included are also the true parameter values, the mean of the estimates, and the proportion of replicates for which the estimated parameter is lower than the true value. Note that all parameters are estimated with reasonable accuracy, except perhaps for the noise level $\sigma_C$. Clearly, the method has some difficulty distinguishing noise on the prey dynamics from noise on the consumption process. This is perhaps not surprising, given that no predators or feeding events are included in the data set.

\section{Discussion}
\label{sec:discussion}

We have advocated the use of state space models based on stochastic differential equations with noisy discrete-time observations, as a basis for analysis of ecological time series, and we have demonstrated that the Laplace approximation is useful for estimating states and parameters in these models. The framework allows for multivariate nonlinear dynamics which represent ecological mechanisms, noise which is not additive, and partial discrete-time observations at irregular sampling times subject to general distributions of measurement noise. We emphasize that the method does not approximate the transition density $p(0,t,x,\cdot)$ with a Gaussian, in contrast to the weak noise expansion which lies beneath extended Kalman filters. The approach leads to compact code which can easily be modified to accommodate changes in model structure, such as changes in drift function, noise intensity, measurement noise distribution and number of state variables. This flexibility is a strong argument for the framework, and is most clearly visible in the code \citep{Thygesen_SDETMB} that accompanies this paper. Such flexibility is particularly important in analysis of ecological time series, where model structures are often determined iteratively during the modeling process rather than fixed a priori. 

We have pointed out that the choice of variables is critical, when the Laplace approximation is formed. Specifically, a substantial bias can appear, if the noise is not additive, the states are root random variables, and the Laplace approximation is based on the mode of their joint posterior distribution. To our knowledge, this phenomenon has not been reported or discussed previously. We have demonstrated that this bias can be eliminated by taking as root random variables the increments in the Brownian motion, which leads to two computational approaches: One where both states and Brownian increments are root variables, and one where the Brownian increments are eliminated and only states remain in the computations. While the latter method is accurate, perhaps more elegant, and lends itself most easily to theoretical analysis, the former is simple and flexible and particularly convenient when parameters must also be estimated, even if it introduces an extra algorithmic parameter ($\epsilon$).

In our continuous-time setting, this bias must be eliminated to obtain consistency when the number of computational time points between observations is increased. However, the discussion has relevance also for fixed computational time steps, i.e., discrete-time models. This leads to the general question in which coordinate system a Laplace approximation should be performed - or, equivalently, how the integrand should be factorized (remark \ref{remark:ambiguity}). It appears that a complete and operational answer to this general question has not yet been given.  

It is important that our proposed method is not restricted to additive noise, since many models in ecology do not have this simplifying structure. It is common to first Lamperti transform the system \citep{Iacus2008}, so that in the new coordinates, the noise is additive. While Lamperti transforms remain an important part of the toolbox, there are three reasons why we prefer to have alternatives: First, the Lamperti transform simplifies the noise structure at the cost of making the drift more complex; this can be cumbersome in early explorative phases of the modeling process, where different forms of drift and noise are tried. Next, the Laplace approximation is typically more accurate in some coordinate systems than others, and this cannot be exploited if the coordinate system is fixed by the requirement of additive noise. For example, for the Cox-Ingersoll-Ross process, the Lamperti transform is given by the square root, but sometimes one may prefer to log-transform, given that all original variables are positive. Finally, many multivariate systems cannot be Lamperti transformed \citep{Moller2010}; this class includes the predator-prey system in section \ref{sec:R-MacA}. 

Conceptually, we believe that it is attractive to use the common machinery of the Laplace approximation to address both transition probabilities between measurements, and the posterior uncertainty on the states at the points of measurements. This facilitates both implementations and analysis of algorithmic performance, and leads to very similar implementations whether the states are observed partially or completely and with or without measurement error. While some inference methods for stochastic differential equations rely on accurate and complete state measurements, the option of including measurement noise is important, in particular as this term may represent not just errors in the measurement process, but also fast unmodeled dynamics \citep{Ljung1999}. We note that also population superstructures can be included, for example, when time series come from a number of animals which differ in specific parameter values; a situation which we explore in ongoing works, where the Laplace approximation serves as a unifying computational tool. 

The question of accuracy of the Laplace approximation remains and will be the topic of further studies. In a specific situation, this accuracy can be tested a posteriori with general methods \citep{Kristensen2016}. It is easy to conceive situations where the approximation will be unsuitable, for example, because there is not a unique optimal path, or because there are locally optimal paths which contribute significantly to the overall probability. Such situations can arise, for example, when there are multiple steady states and near regime shifts \citep{Medeiros2025,Thygesen2025regimes}, or in oscillatory systems such as the predator-prey system of section \ref{sec:R-MacA}, where paths with different number of cycles but same endpoints can coexist and connect points of observations. These examples suggest that the Laplace approximation require not too slow sampling for such systems; note that very slow sampling usually prohibits inference of dynamics anyway. 

The accuracy of the algorithm depends on the number of intermediate time points inserted between times of observations. This trade-off between computing times and accuracy must at present be done partially ad hoc, although support can come the accuracy of the time discretization employed, as in section \ref{sec:an-example:-cox}. The combination of different discretization methods (for example, Strang splitting \citep{Pilipovic2024})  and the Laplace approximation is a topic of current work and will be reported elsewhere  \citep{Thygesen2025sdeB}, including the continuous-time limit of infinitely many intermediate time steps. 

Our method yields posterior estimates of the states, based on the entire time series of data points, both past and future. That is, in the linear-Gaussian case, we reproduce the \emph{smoothing}\/ Kalman filter, even if it is straightforward to include unobserved states after the last observation to obtain forecasting. On consequence of this is that the residuals from the model are unsuitable for model validation. As discussed by \cite{Thygesen2017}, one-step predictions can be formed by iteratively including data points, but computationally, the more attractive technique may be to sample a single realization of the driving Brownian motion from the posterior distribution, and base model validation on patterns in this realization.

Regarding the choice between the \Ito\ and the Stratonovich interpretation of the noise, it is convenient that Laplace approximation apply equally well in both cases. The choice of interpretation does have some effect on the most probable path, around which the Laplace approximation is performed, even if one does the standard conversion between the drift terms in the \Ito\ and Stratonivich equations. Although the difference appears to be minor in numerical examples, the two interpretations lead to slightly different approximations of the transition probabilities for the same diffusion process.  

The ambition of this paper has been to present the approach and demonstrate its feasibility, while a number of issues remain topics of future studies. These include the continuous-time limit \citep{Thygesen2025sdeB} which provides stronger theoretical support, assists with error analysis, and facilitates studying different discretization methods. Although the method can be applied to general non-linear stochastic differential equations with arbitrary models of measurement noise, it is difficult to assess the accuracy of and potential bias in parameter estimates; therefore, simulation-reestimation experiments remain important, as in section \ref{sec:R-MacA}.  Regardless of these outstanding issues, we believe that the method's flexibility, broad applicability and ease of implementation make it a valuable supplement to existing techniques. 

\section*{Ethics declaration}
\label{sec:ethics-declaration}

\subsection*{Conflicts of interest}
\label{sec:conflicts-interest}

The authors declare no competing interests.

\appendix

\section{Example implementation of the Laplace approximation}
\label{sec:code}

The following \texttt{R} code illustrates how to evaluate the transition probability given by a stochastic differential equation, inserting intermediate points between the start and end time. The function evaluates the joint density of all random variables. 

\begin{verbatim}
## TMB: Ito formulation, dB and X as root variables
nloglik.XdB <- function(p)
{
    dX <- diff(p$X)               # Compute state increments
    xX <- head(p$X,-1)            # Old states
    XY <- p$X[c(1,length(p$X))]   # Start and end point

    ## Euler-Maruyama relationship 
    dXpred <- fI(xX)*p$dt + g(xX)*p$dB

    nll <- - sum(dnorm(p$dB,0,sqrt(p$dt),log=TRUE))
    nll <- nll - sum(dnorm(dX,dXpred,p$epsilon*sqrt(p$dt),log=TRUE))
    nll <- nll - sum(dnorm(XY,c(p$x,p$y),p$epsilon,log=TRUE))

    return(nll)
}
\end{verbatim}

The three contributions to the negative log-likelihood are that the increments in the Brownian motion should be Gaussian, and that the Euler-Maruyama relationship should hold approximately, as should the start and end points. From this user code, the steps of computing the Laplace approximation are automated by the \texttt{RTMB} package \cite{RTMB}. See \citep{Thygesen_SDETMB}  for full code.

\section{Laplace approximation in the Stratonovich formulation}
\label{app:Strat}

Here, we detail the Laplace approximation of the transition densities for the Stratonovich equation \eqref{eq:Strat-SDE} based on the discretization scheme \eqref{eq:Strat-approximation}.  First, condition on $X_{t}=x$ in this scheme and use the shorthand $\Delta B = B_{t+h} - B_t$, then we can rewrite the scheme  as
\[
  \eta(X_{t+h},\Delta B) = 0
\]
where the function $\eta:\RR^n \times \RR^n \mapsto \RR^n$ is given by 
\[
  \eta(y,b) = y - x - \frac 12 (\fS(x)+\fS(y)) h - \frac 12 (g(x) + g(y)) b .
\]
This defines, at least locally and generically, a diffeomorphism between the new state $y$ and the Brownian increment $b$, and therefore allows us to express the p.d.f. of $X_{t+h}$ at $y$ in terms of the p.d.f. of $\Delta B$ at $b$. Using $\phi_{\Delta B}(b)$ and $\phi_{X_{t+h}}(y)$ for these p.d.f.'s, we have the relationship
\[
  \left| \delvis \eta b(y,b) \right| ^{-1} \phi_{ \Delta B}(b) =
  \left| \delvis \eta y (y,b) \right| ^{-1} \phi_{ X_{t+h}}(y) 
\]
assuming that the map between $y$ and $b$ is in fact one-to-one. Here, we have
\[
  \delvis \eta y = I - \frac h2 \nabla f - \sum_k \frac {b_k} 2  \nabla g_k,
  \quad \delvis \eta b = - \frac{g(x)+g(y)}2 . 
\]
Since the increment $\Delta B$ is distributed as a Gaussian, $N(0,Ih)$ where $I$ is the $n$-dimensional identity matrix, this leads to the approximate transition probability density:
\begin{multline}
  \hat p(0,x,h,y,h) = \\ \left| I - \frac h2 \nabla \fS(y) - \sum_k \frac {b_k}2  \nabla g_k(y)  \right| \cdot \left| \frac{g(x) + g(y)}2\right|^{-1}  | 2 \pi h|^{-n/2} e^{- |b|^2/(2h)} . 
\end{multline}
Here, $b=(b_1,\ldots,b_n)\in \RR^n$ is the solution to the equation system $\eta(y,b)=0$, i.e.,
\begin{equation}
  \label{eq:b-solve}
  b = (g(x) + g(y))^{-1} (2 y - 2 x  - (\fS(x ) + \fS(y)) h) . 
\end{equation}
where we assume the inverse is well defined, which seems plausible given that $g(x)$ and $g(y)$ are invertible by assumption and that $x$ will be near $y$. 

We can now construct the Laplace approximation $\hat p(0,x,t,y,h)$ of the transition density $p(0,x,t,y)$ using the same principle as for the \Ito\ equation, i.e., as in section \ref{sec:form-state-space}. As latent variables we use $\bar x = (x_1,\ldots,x_{N-1})$ where $x_i\in \RR^n$ is the state vector at an intermediate time point $t_i = ih$; here, $Nh=t$. From these, we can compute the increments $b_i\in \RR^n$ of the Brownian motion over each time interval $[t_{i-1},t_i]$ from~\eqref{eq:b-solve}, yielding
\begin{equation}
  \label{eq:Strat-dB}
  b_i = (g(x_i) + g(x_{i-1}))^{-1} (2 x_i - 2 x_{i-1}  - (\fS(x_{i-1} ) + \fS(x_i)) h) . 
\end{equation}
Following section \ref{sec:form-state-space}, we form the joint log-density of the increments
\[
  \gamma(\bar x) = - \sum_{i=1}^N [ \log \phi( b_i / \sqrt{ \Delta t_i})  - \frac n2 \log \Delta t_i ] 
\]
and maximize this w.r.t. $\bar x$, using the symbol $\bar x^*$ for the maximum point. Then the Laplace approximation is
\begin{multline}
  \label{eq:Laplace-Stratonovich}
  \hat p(0,x,t,y) = \\
  | H / (2 \pi) |^{-1/2} \exp(-\gamma(\bar x^*)) \cdot \prod_{i=0}^{N-1} \frac{\left| I - \frac h2 \nabla \fS(x^*_{i+1})  - \sum_k \frac {b^*_{ki}}2  \nabla g_k (x^*_{i+1})\right| }
  {\left|\frac 12 (g(x^*_i) + g(x^*_{i+1}))\right|}
\end{multline}
where $H$ is the Hessian of $\psi$ at $\bar x^*$, and $b^*_{ki}$ is the corresponding increment of the $k$'th component of the Brownian motion over the $i$'th time step. As in section \ref{sec:form-state-space}, we use the automation in \texttt{RTMB} to address the first two factors, and then correct the result with the last product. See \citep{Thygesen_SDETMB} for the full implementation.

\bibliographystyle{plainnat}
\bibliography{new}

\begin{thebibliography}{46}
\providecommand{\natexlab}[1]{#1}
\providecommand{\url}[1]{\texttt{#1}}
\expandafter\ifx\csname urlstyle\endcsname\relax
  \providecommand{\doi}[1]{doi: #1}\else
  \providecommand{\doi}{doi: \begingroup \urlstyle{rm}\Url}\fi

\bibitem[A{\"\i}t-Sahalia(2002)]{AitSahalia2002}
Yacine A{\"\i}t-Sahalia.
\newblock Maximum likelihood estimation of discretely sampled diffusions: {A}
  closed-form approximation approach.
\newblock \emph{Econometrica}, 70\penalty0 (1):\penalty0 223--262, 2002.

\bibitem[Albertsen et~al.(2015)Albertsen, Whoriskey, Yurkowski, Nielsen, and
  Flemming]{Albertsen2015}
C.M. Albertsen, K.~Whoriskey, D.~Yurkowski, A.~Nielsen, and J.M. Flemming.
\newblock Fast fitting of non-{Gaussian} state-space models to animal movement
  data via {Template Model Builder}.
\newblock \emph{Ecology}, 96\penalty0 (10):\penalty0 2598--2604, 2015.

\bibitem[Auger-Méthé et~al.(2021)Auger-Méthé, Newman, Cole, Empacher,
  Gryba, King, Leos-Barajas, Mills~Flemming, Nielsen, Petris, and
  Thomas]{Auger-Methe2021}
Marie Auger-Méthé, Ken Newman, Diana Cole, Fanny Empacher, Rowenna Gryba,
  Aaron~A. King, Vianey Leos-Barajas, Joanna Mills~Flemming, Anders Nielsen,
  Giovanni Petris, and Len Thomas.
\newblock A guide to state–space modeling of ecological time series.
\newblock \emph{Ecological Monographs}, 91\penalty0 (4):\penalty0 e01470, 2021.
\newblock \doi{https://doi.org/10.1002/ecm.1470}.
\newblock URL
  \url{https://esajournals.onlinelibrary.wiley.com/doi/abs/10.1002/ecm.1470}.

\bibitem[Braumann(2007)]{Braumann2007}
Carlos~A. Braumann.
\newblock It{\^o} versus stratonovich calculus in random population growth.
\newblock \emph{Mathematical Biosciences}, 206\penalty0 (1):\penalty0 81--107,
  2007.
\newblock ISSN 0025-5564.
\newblock \doi{https://doi.org/10.1016/j.mbs.2004.09.002}.
\newblock URL
  \url{https://www.sciencedirect.com/science/article/pii/S0025556405001537}.
\newblock Alcala Special Issue.

\bibitem[Brooks et~al.(2017)Brooks, Kristensen, {van Benthem}, Magnusson, Berg,
  Nielsen, Skaug, Maechler, and Bolker]{Brooks2017}
Mollie~E. Brooks, Kasper Kristensen, Koen~J. {van Benthem}, Arni Magnusson,
  Casper~W. Berg, Anders Nielsen, Hans~J. Skaug, Martin Maechler, and
  Benjamin~M. Bolker.
\newblock {glmmTMB} balances speed and flexibility among packages for
  zero-inflated generalized linear mixed modeling.
\newblock \emph{The R Journal}, 9\penalty0 (2):\penalty0 378--400, 2017.
\newblock \doi{10.32614/RJ-2017-066}.

\bibitem[Cox et~al.(1985)Cox, Ingersoll~Jr, and Ross]{Cox1985}
J.C. Cox, J.E. Ingersoll~Jr, and S.A. Ross.
\newblock {A theory of the term structure of interest rates}.
\newblock \emph{Econometrica: Journal of the Econometric Society}, pages
  385--407, 1985.

\bibitem[Evensen(2003)]{Evensen2003}
G.~Evensen.
\newblock The ensemble {Kalman} filter: Theoretical formulation and practical
  implementation.
\newblock \emph{Ocean dynamics}, 53\penalty0 (4):\penalty0 343--367, 2003.

\bibitem[Fournier et~al.(2012)Fournier, Skaug, Ancheta, Ianelli, Magnusson,
  Maunder, Nielsen, and Sibert]{Fournier2012}
David~A. Fournier, Hans~J. Skaug, Johnoel Ancheta, James Ianelli, Arni
  Magnusson, Mark~N. Maunder, Anders Nielsen, and John Sibert.
\newblock Ad model builder: using automatic differentiation for statistical
  inference of highly parameterized complex nonlinear models.
\newblock \emph{Optimization Methods and Software}, 27\penalty0 (2):\penalty0
  233--249, 2012.
\newblock \doi{10.1080/10556788.2011.597854}.
\newblock URL \url{https://doi.org/10.1080/10556788.2011.597854}.

\bibitem[Fuchs(2013)]{Fuchs2013}
Christiane Fuchs.
\newblock \emph{Inference for Diffusion Processes}.
\newblock Springer, Berlin, Heidelberg, 2013.

\bibitem[Gordon et~al.(1993)Gordon, Salmond, and Smith]{Gordon1993}
Neil~J Gordon, David~J Salmond, and Adrian~FM Smith.
\newblock Novel approach to nonlinear/non-{Gaussian} {Bayesian} state
  estimation.
\newblock In \emph{IEE proceedings F (radar and signal processing)}, volume
  140, pages 107--113. IET, 1993.

\bibitem[Holst et~al.(2003)Holst, Lindström, Madsen, and {Aalborg
  Nielsen}]{Holst2003}
Jan Holst, Erik Lindström, Henrik Madsen, and Henrik {Aalborg Nielsen}.
\newblock Model validation in non-linear continuous-discrete grey-box models.
\newblock \emph{IFAC Proceedings Volumes}, 36\penalty0 (16):\penalty0
  1495--1500, 2003.
\newblock ISSN 1474-6670.
\newblock \doi{https://doi.org/10.1016/S1474-6670(17)34971-6}.
\newblock URL
  \url{https://www.sciencedirect.com/science/article/pii/S1474667017349716}.
\newblock 13th IFAC Symposium on System Identification (SYSID 2003), Rotterdam,
  The Netherlands, 27-29 August, 2003.

\bibitem[Iacus(2008)]{Iacus2008}
S.M. Iacus.
\newblock \emph{Simulation and inference for stochastic differential equations:
  with {R} examples}.
\newblock Springer Verlag, New York, 2008.

\bibitem[Jamba et~al.(2024)Jamba, Jacinto, Filipe, and Braumann]{Jamba2024}
Nelson~T. Jamba, Gonçalo Jacinto, Patrícia~A. Filipe, and Carlos~A. Braumann.
\newblock Estimation for stochastic differential equation mixed models using
  approximation methods.
\newblock \emph{AIMS Mathematics}, 9\penalty0 (4):\penalty0 7866--7894, 2024.
\newblock \doi{10.3934/math.2024383}.

\bibitem[Kalman and Bucy(1961)]{Kalman1961}
R.E. Kalman and R.S. Bucy.
\newblock New results in linear filtering and prediction theory.
\newblock \emph{Journal of Basic Engineering}, pages 95--108, 1961.

\bibitem[Karimi and McAuley(2014)]{Karimi2014}
Hadiseh Karimi and Kimberley~B. McAuley.
\newblock A maximum-likelihood method for estimating parameters, stochastic
  disturbance intensities and measurement noise variances in nonlinear dynamic
  models with process disturbances.
\newblock \emph{Computers \& Chemical Engineering}, 67:\penalty0 178--198,
  2014.
\newblock ISSN 0098-1354.
\newblock \doi{https://doi.org/10.1016/j.compchemeng.2014.04.007}.
\newblock URL
  \url{https://www.sciencedirect.com/science/article/pii/S0098135414001240}.

\bibitem[Karimi and McAuley(2016)]{Karimi2016}
Hadiseh Karimi and Kimberley~B. McAuley.
\newblock Bayesian estimation in stochastic differential equation models via
  {Laplace} approximation.
\newblock \emph{IFAC-PapersOnLine}, 49\penalty0 (7):\penalty0 1109--1114, 2016.
\newblock ISSN 2405-8963.
\newblock \doi{https://doi.org/10.1016/j.ifacol.2016.07.351}.
\newblock URL
  \url{https://www.sciencedirect.com/science/article/pii/S2405896316305584}.
\newblock 11th IFAC Symposium on Dynamics and Control of Process Systems
  Including Biosystems DYCOPS-CAB 2016.

\bibitem[Kessler(1997)]{Kessler1997}
Mathieu Kessler.
\newblock Estimation of an ergodic diffusion from discrete observations.
\newblock \emph{Scandinavian Journal of Statistics}, 24:\penalty0 211--229,
  1997.
\newblock \doi{10.1111/1467-9469.00059}.

\bibitem[Kloeden and Platen(1999)]{Kloeden1999}
P.E. Kloeden and E.~Platen.
\newblock \emph{Numerical Solution of Stochastic Differential Equations}.
\newblock Springer, New York, third edition, 1999.

\bibitem[Krainski et~al.(2019)Krainski, Gómez-Rubio, Bakka, Lenzi,
  Castro-Camilo, Simpson, Lindgren, and Rue]{Krainski2019}
Elias Krainski, Virgilio Gómez-Rubio, Haakon Bakka, Amanda Lenzi, Daniela
  Castro-Camilo, Daniel Simpson, Finn Lindgren, and Håvard Rue.
\newblock \emph{Advanced Spatial Modeling with Stochastic Partial Differential
  Equations Using {R} and {INLA}}.
\newblock CRC Press, London, 2019.

\bibitem[Kristensen(2025)]{RTMB}
Kasper Kristensen.
\newblock \emph{RTMB: 'R' Bindings for 'TMB'}, 2025.
\newblock URL \url{https://CRAN.R-project.org/package=RTMB}.
\newblock R package version 1.7.

\bibitem[Kristensen et~al.(2016)Kristensen, Nielsen, Berg, Skaug, and
  Bell]{Kristensen2016}
Kasper Kristensen, Anders Nielsen, Casper~W. Berg, Hans Skaug, and Bradley~M.
  Bell.
\newblock {TMB}: Automatic differentiation and {Laplace} approximation.
\newblock \emph{Journal of Statistical Software}, 70, 2016.
\newblock \doi{10.18637/jss.v070.i05}.

\bibitem[Kristensen et~al.(2004)Kristensen, Madsen, and
  J{\o}rgensen]{Kristensen2004}
Niels~Rode Kristensen, Henrik Madsen, and Sten~Bay J{\o}rgensen.
\newblock Parameter estimation in stochastic grey-box models.
\newblock \emph{Automatica}, 40\penalty0 (2):\penalty0 225--237, 2004.

\bibitem[Lavender et~al.(2026)Lavender, Scheidegger, Moor, and
  Albert]{Lavender2025}
Edward Lavender, Andreas Scheidegger, Helen Moor, and Carlo Albert.
\newblock State-space models and inference approaches for aquatic animal
  tracking with passive acoustic telemetry and biologging sensors.
\newblock \emph{Methods in Ecology and Evolution}, 17\penalty0 (2):\penalty0
  418--434, 2026.
\newblock \doi{https://doi.org/10.1111/2041-210x.70186}.
\newblock URL
  \url{https://besjournals.onlinelibrary.wiley.com/doi/abs/10.1111/2041-210x.70186}.

\bibitem[Ljung(1999)]{Ljung1999}
L.~Ljung.
\newblock \emph{System Identification - Theory for the User}.
\newblock Information and System Sciences Series. Prentice-Hall, second
  edition, 1999.

\bibitem[Madsen(2007)]{Madsen2007}
H.~Madsen.
\newblock \emph{Time series analysis}.
\newblock Chapman \& Hall/CRC, London, 2007.

\bibitem[Markussen(2009)]{Markussen2009}
Bo~Markussen.
\newblock Laplace approximation of transition densities posed as {Brownian}
  expectations.
\newblock \emph{Stochastic Processes and their Applications}, 119:\penalty0
  208--231, 2009.

\bibitem[Medeiros et~al.(2025)Medeiros, Sorenson, Johnson, Palkovacs, and
  Munch]{Medeiros2025}
Lucas~P Medeiros, Darian~K Sorenson, Bethany~J Johnson, Eric~P Palkovacs, and
  Stephan~B Munch.
\newblock Revealing unseen dynamical regimes of ecosystems from population
  time-series data.
\newblock \emph{Proceedings of the National Academy of Sciences}, 122\penalty0
  (24):\penalty0 e2416637122, 2025.

\bibitem[M{\o}ller and Madsen(2010)]{Moller2010}
Jan~Kloppenborg M{\o}ller and Henrik Madsen.
\newblock From state dependent diffusion to constant diffusion in stochastic
  differential equations by the {Lamperti} transform.
\newblock Technical Report IMM-Technical Report-2010-16, 2010.

\bibitem[Murray(1989)]{Murray1989}
J.D. Murray.
\newblock \emph{Mathematical Biology}.
\newblock Springer-Verlag, New York, 1989.

\bibitem[Nabeel et~al.(2025)Nabeel, Karichannavar, Palathingal, Jhawar,
  Br\"{u}ckner, Raj~M, and Guttal]{Nabeel2025}
Arshed Nabeel, Ashwin Karichannavar, Shuaib Palathingal, Jitesh Jhawar,
  David~B. Br\"{u}ckner, Danny Raj~M, and Vishwesha Guttal.
\newblock Discovering stochastic dynamical equations from ecological time
  series data.
\newblock \emph{The American Naturalist}, 205\penalty0 (4):\penalty0
  E100--E117, 2025.
\newblock \doi{10.1086/734083}.
\newblock URL \url{https://doi.org/10.1086/734083}.
\newblock PMID: 40179429.

\bibitem[Nielsen and Berg(2014)]{Nielsen2016SAM}
Anders Nielsen and Casper~W. Berg.
\newblock Estimation of time-varying selectivity in stock assessments using
  state-space models.
\newblock \emph{Fisheries Research}, 158:\penalty0 96--101, 2014.
\newblock ISSN 0165-7836.
\newblock \doi{https://doi.org/10.1016/j.fishres.2014.01.014}.
\newblock URL
  \url{https://www.sciencedirect.com/science/article/pii/S0165783614000228}.
\newblock SI: Selectivity.

\bibitem[Nielsen et~al.(2000)Nielsen, Madsen, and Young]{Nielsen2000}
Jan~Nygaard Nielsen, Henrik Madsen, and Peter~C. Young.
\newblock Parameter estimation in stochastic differential equations: An
  overview.
\newblock \emph{Annual Reviews in Control}, 24:\penalty0 83--94, 2000.
\newblock ISSN 1367-5788.
\newblock \doi{https://doi.org/10.1016/S1367-5788(00)90017-8}.
\newblock URL
  \url{https://www.sciencedirect.com/science/article/pii/S1367578800900178}.

\bibitem[{\O}ksendal(2010)]{Oeksendal2010}
B.~{\O}ksendal.
\newblock \emph{Stochastic Differential Equations - An Introduction with
  Applications}.
\newblock Springer-Verlag, sixth edition, 2010.

\bibitem[Pedersen et~al.(2008)Pedersen, Righton, Thygesen, Andersen, and
  Madsen]{Pedersen2008}
M.W. Pedersen, D.~Righton, U.H. Thygesen, K.H. Andersen, and H.~Madsen.
\newblock {Geolocation of {North Sea} cod (\emph{Gadus morhua}) using hidden
  {Markov} models and behavioural switching}.
\newblock \emph{Canadian Journal of Fisheries and Aquatic Sciences},
  65\penalty0 (11):\penalty0 2367--2377, 2008.

\bibitem[Pilipovic et~al.(2024)Pilipovic, Samson, and Ditlevsen]{Pilipovic2024}
Predrag Pilipovic, Adeline Samson, and Susanne Ditlevsen.
\newblock Parameter estimation in nonlinear multivariate stochastic
  differential equations based on splitting schemes.
\newblock \emph{The Annals of Statistics}, 52\penalty0 (2):\penalty0 842--867,
  2024.

\bibitem[{Prakasa Rao} and Rubin(1979)]{PrakasaRao1979}
B.L.S. {Prakasa Rao} and H.~Rubin.
\newblock Asymptotic theory for process least squares estimators for diffusion
  processes.
\newblock Technical Report 79-13, Department of Statistics, Purdue University,
  1979.

\bibitem[Rosenzweig and MacArthur(1963)]{Rosenzweig1963}
Michael~L Rosenzweig and Robert~H MacArthur.
\newblock Graphical representation and stability conditions of predator-prey
  interactions.
\newblock \emph{The American Naturalist}, 97\penalty0 (895):\penalty0 209--223,
  1963.

\bibitem[Rue et~al.(2009)Rue, Martino, and Chopin]{Rue2009}
Håvard Rue, Sara Martino, and Nicolas Chopin.
\newblock Approximate bayesian inference for latent gaussian models by using
  integrated nested laplace approximations.
\newblock \emph{Journal of the Royal Statistical Society Series B: Statistical
  Methodology}, 71\penalty0 (2):\penalty0 319--392, 04 2009.

\bibitem[Simon(2006)]{Simon2006}
D.~Simon.
\newblock \emph{Optimal State Estimation - {Kalman}, {$\cal H_\infty$}, and
  Nonlinear Approaches}.
\newblock John Wiley \& Sons, Hoboken, New Jersey, 2006.

\bibitem[Thygesen(2023)]{Thygesen2023sde}
Uffe~H{\o}gsbro Thygesen.
\newblock \emph{Stochastic Differential Equations for Science and Engineering}.
\newblock Chapman \& Hall/CRC, New York, 2023.
\newblock \doi{10.1201/9781003277569}.

\bibitem[Thygesen et~al.(2017)Thygesen, Albertsen, Berg, Kristensen, and
  Nielsen]{Thygesen2017}
Uffe~H{\o}gsbro Thygesen, Christoffer~Moesgaard Albertsen, Casper~Willestofte
  Berg, Kasper Kristensen, and Anders Nielsen.
\newblock Validation of ecological state space models using the {Laplace}
  approximation.
\newblock \emph{Environmental and Ecological Statistics}, 24\penalty0
  (2):\penalty0 317--339, 2017.
\newblock ISSN 1573-3009.
\newblock \doi{10.1007/s10651-017-0372-4}.
\newblock URL \url{http://dx.doi.org/10.1007/s10651-017-0372-4}.

\bibitem[Thygesen(2025{\natexlab{a}})]{Thygesen2025regimes}
Uffe~Høgsbro Thygesen.
\newblock Predicting regime shifts and beyond.
\newblock \emph{Proceedings of the National Academy of Sciences}, 122\penalty0
  (29):\penalty0 e2513605122, 2025{\natexlab{a}}.
\newblock \doi{10.1073/pnas.2513605122}.
\newblock URL \url{https://www.pnas.org/doi/abs/10.1073/pnas.2513605122}.

\bibitem[Thygesen(2025{\natexlab{b}})]{Thygesen2025sdeB}
Uffe~Høgsbro Thygesen.
\newblock Transition probabilities for stochastic differential equations using
  the {Laplace} approximation: Analysis of the continuous-time limit.
\newblock \emph{arXiv}, page 2503.21399, 2025{\natexlab{b}}.
\newblock URL \url{https://arxiv.org/abs/2503.21399}.
\newblock https://doi.org/10.48550/arXiv.2503.21399.

\bibitem[Thygesen(2025{\natexlab{c}})]{Thygesen_SDETMB}
Uffe~Høgsbro Thygesen.
\newblock {SDE-TMB}, 2025{\natexlab{c}}.
\newblock URL \url{https://github.com/Uffe-H-Thygesen/SDE-TMB}.
\newblock R code available at github.com/Uffe-H-Thygesen/SDE-TMB.

\bibitem[Thygesen(2025{\natexlab{d}})]{Thygesen_SDEtools}
Uffe~Høgsbro Thygesen.
\newblock {SDEtools}, 2025{\natexlab{d}}.
\newblock URL \url{https://github.com/Uffe-H-Thygesen/SDEtools}.
\newblock R package available at github.com/Uffe-H-Thygesen/SDEtools.

\bibitem[Zucchini and MacDonald(2009)]{Zucchini2009}
Walter Zucchini and Iain~L MacDonald.
\newblock \emph{Hidden Markov models for time series: an introduction using R}.
\newblock CRC Press, London, 2009.

\end{thebibliography}

\end{document}